# Automated manipulation of free space optical beams with integrated silicon photonic meshes

**Maziyar Milanizadeh[1], Fabio Toso[1], Giorgio Ferrari[1], Tigers Jonuzi[1,+],
David A. B. Miller[2], Andrea Melloni[1], Francesco Morichetti[1,*]**

[1]*Dipartimento di Elettronica, Informazione e Bioingegneria - Politecnico di Milano, Milano, 20133 Italy*
[2]*Ginzton Laboratory, Stanford University, Spilker Building, Stanford, CA 94305, USA*
*\*Corresponding author: francesco.morichetti@polimi.it*
[+]*now with VLC Photonics, Universidad Politécnica de Valencia, 46022 Valencia, Spain*

**Free-space optics (FSO) offer promising solutions in wireless systems where many different devices need to communicate and interoperate, exchanging massive amounts of data as in communication and sensor networks, and in technologies for positioning and ranging. In these applications, control and manipulation of the light field is required for properly generating and correctly receiving FSO beams even in the presence of unpredictable objects and turbulence in the light path. In this work, we use a programmable mesh of Mach-Zehnder Interferometers (MZIs) to automatically control the complex field radiated and captured by an array of optical antennas. Several functionalities are demonstrated, including the generation of perfectly shaped beams with non-perfect optical antennas, the imaging of a desired field pattern through an obstacle or a diffusive medium, and the identification of an unknown obstacle inserted in the FSO path. Compared to conventional devices used for the manipulation of FSO beams, such as spatial light modulators (SLMs), our programmable circuit can self-configure through automated control strategies and can be integrated with other functionalities implemented onto the same photonic chip.**

## 1. INTRODUCTION

The possibility of manipulating the properties of a light beam is a fundamental requirement of free-space optical (FSO) systems. Structured light [1] – that is, optical beams with optimized amplitude, phase and polarization profiles – is exploited in high-resolution imaging [2], in microscopy [3], in particle detection, localization and manipulation [4], and in communications [5]. In these applications FSO beams are conventionally generated, manipulated and received by using bulk optics, such as classical lenses or spatial light modulators (SLMs). SLMs allow the independent control of amplitude and the phase in a large number of points (pixels) in the cross-sectional plane of the beam, as in iquid-crystal on silicon (LCOS) technology [6], but at the price of quite large size and cost, and loss in any amplitude modulation. However, massive connection of devices driven by the internet of things (IoT) asks for solutions where footprint, weight and cost reduction are of primary importance, for instance, in systems for localization, positioning and ranging for autonomous vehicles and drones, as well as artificial vision and imaging systems for portable devices. More compact SLM architectures have been recently proposed, which exploit all-solid state tunable metasurfaces [7], [8]. Although allowing a significant device downscaling, these devices are hardly integrable with other functionalities, like tuneable wavelength selective filters, splitters/combiners and (de)multiplexers, nor with fast time-domain modulators, which could allow evolution towards more advanced all-optical processing of FSO beams. Moreover, automated configuration of SLM systems require interferometric phase calibration techniques and optimization algorithms, whose computational complexity scales heavily with the pixel count [9].

In this scenario, programmable photonic circuits represent a promising approach for the advanced processing of the light on a chip [10]. Programmable photonic circuits consist of generic reconfigurable architectures, with either a feed-forward [11] [12] [13] [14] or recursive [15] [16] mesh topology, that can be reconfigured during operation to implement arbitrary linear functions. Due to the inherent general-purpose nature of these circuits, many applications have been proposed in different fields, including microwave photonics [17] [18], manipulation and unscrambling of guided modes for telecommunications [13] [19] [20], vector-matrix multiplication and computing [21], quantum information processing [12] [14] [22] and artificial neural networks [14] [23].

In this work, we demonstrate the manipulation of FSO beams on a silicon chip using a programmable mesh of Mach-Zehnder Interferometers (MZIs). By automatically controlling the complex field radiated and captured by an array of optical antennas, a number of functionalities are demonstrated with such a mesh; these functionalities go well beyond the capabilities of ordinary optical phased arrays. The architecture of the photonic circuit is described in Sec. 2. The I/O interface between the light guided in the chip and the FSO beams is provided by vertically emitting grating couplers used as optical antennas. In Sec. 3, we show that, since the amplitude and the phase of the light can be individually controlled at each antenna, perfect beams with an ideal pattern can be generated from such imperfect optical antennas. Then we demonstrate the possibility of shaping a light beam in a such a way that the desired pattern is observed at the receiver, even in the presence of an obstacle (phase mask or diffusive medium) inserted in the light path (Sec. 4). This implies also the possibility of automatically establishing optimum free-space communication channels, a concept theoretically proposed a few years ago [24], but never demonstrated so far. Such functionality also means we are effectively performing phase conjugation with this system. Finally, we use the proposed device for the identification of an unknown obstacle inserted in the FSO path.

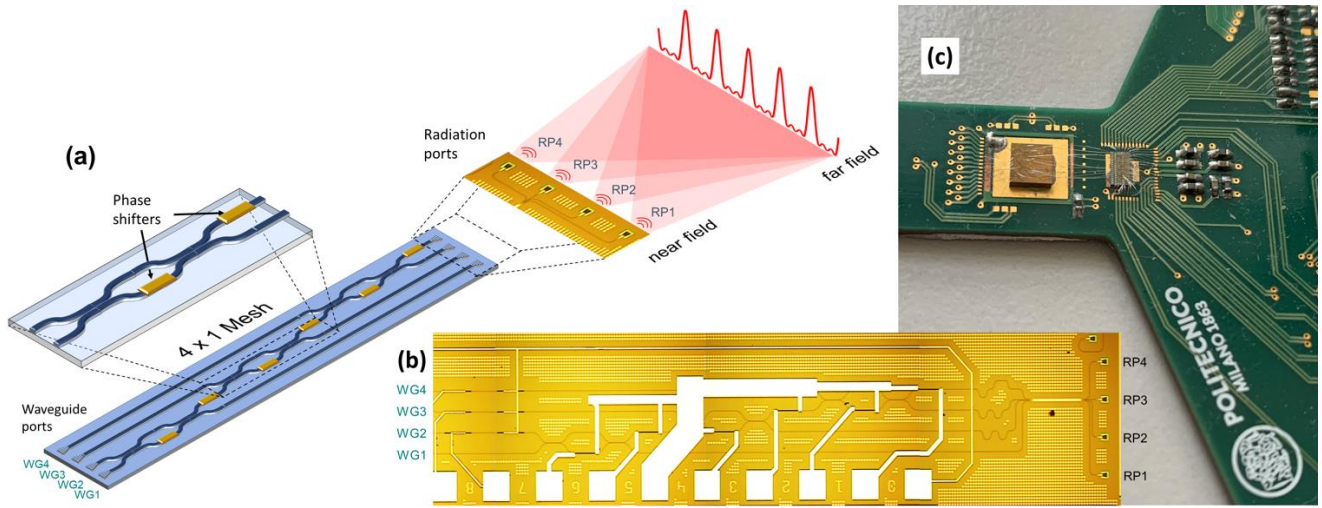

**Fig. 1.** (a) Schematic representation of a 4 x 1 "diagonal line" mesh and (b) its implementation on a conventional silicon photonic platform. The overall footprint of the circuit is 3.7 mm by 1.5mm and is mounted on a PCB with a CMOS ASIC for the readout of the on-chip sensors and driving the integrated thermal tuners.

## 2. Programmable mesh architecture

The programmable photonic circuit employed in this work consists of a diagonal mesh of tunable beam splitters, which are arranged as in the scheme of Fig. 1a. Each tunable beam splitter is implemented through a balanced MZI with two integrated phase shifters enabling the control of the amplitude split ratio and of the relative phase shift between the optical fields at the output ports. To use the mesh as a generator (transmitter) of FSO beams, the light is coupled into individual waveguides through terminations we call waveguide ports (WGs), while on the other side, terminations we call radiation ports (RPs) are used as optical antennas to radiate the guided light out to free space. The same circuit can be used to receive free-space beams (receiver) and in this case the RPs are used to couple free-space optical beams into the photonic chip. This mesh architecture was previously employed to implement several optical functions with FSO beams, including beam steering, beam identification and coupling of a free space beam coming from an arbitrary direction [25].

As shown in Fig. 1a, the far field pattern of the radiated light is the Fourier transform of the field pattern at the RPs [26]. By tuning the integrated MZIs, both the amplitude and the phase of the light radiated by the RPs can be controlled; thus, the shape and direction of the far field beam can be modified accordingly. The number of RPs identifies the spatial harmonics available for generation of FSO beam shapes, corresponding to the degrees of freedom for beam manipulation. The device concept employed in this work includes four RPs, but the results are scalable to meshes with an arbitrary number of ports.

The photonic circuit was fabricated on a standard 220 nm silicon photonic platform by using 500-nm-wide channel waveguides (see Fig. 1b). The RPs consist of a linear array of grating couplers with a mutual spacing of 127 µm. The beam emitted by each grating coupler has an approximately gaussian profile with a 3-degree beam width. The two beam splitters inside the MZI are 3 dB directional couplers with a gap distance of 300 nm and a length of 40 µm. The phase shifters employ thermal tuners made of TiN metal strips (2 µm x 80 µm) deposited at a distance of 1 µm from the top of the silicon waveguides. The same TiN layer is used to implement on-chip transparent photodetectors (CLIPP [27]) to locally monitor the switching state of each MZI and implement automatic tuning and stabilization procedures. The silicon photonic chip was mounted on the electronic printed circuit board (PCB) shown in Fig. 1(c), which also hosts the drivers for the thermal tuners and the electronics (ASIC) for the read-out of the on-chip detectors. Dithering-based control algorithms relying on a thermal-eigenmode decomposition (TED) method are used to mitigate on-chip thermal crosstalk effects between the phase shifters [28] [29]. More details on the circuit design, fabrication technology and the CMOS ASIC electronic circuit [30] for the automatic control of the mesh can be found in [13]. All the experiments reported in this work were carried out at a wavelength of 1550 nm.

## 3. Perfect beams with non-perfect optical antennas

A remarkable property of our approach is that phase and amplitude imperfections in the plane of the RPs can be automatically compensated by a suitable tuning of the integrated MZIs. This enables generating "perfectly-shaped" free-space beams suitable for the propagating channel even if the behavior of the radiating elements is different from the nominal one.

To prove this concept, we employed the free-space setup shown in Fig. 2(a); this allows us both to (i) intentionally corrupt the field pattern of the RPs, and (ii) to monitor the shape of the far field in order to provide a feedback signal to the system controlling the tuning status of the mesh. The output fields of the RPs are brought to the Fourier plane (P1) through a Fourier transforming biconvex lens (100 mm focal length). An identical lens is used to bring the field back to an image plane (P2), where a replica of the four RPs pattern can be observed (see Supplementary Sec. S1 for details of the experimental setup). A second Fourier-transforming lens (250 mm focal length) transforms the field in plane P2 into plane P3. The field in this plane P3 therefore represents what would be the far field diffraction from the output of plane P2. We then image P3 into the camera by focusing the camera (at a distance of 350 mm) onto plane P3.

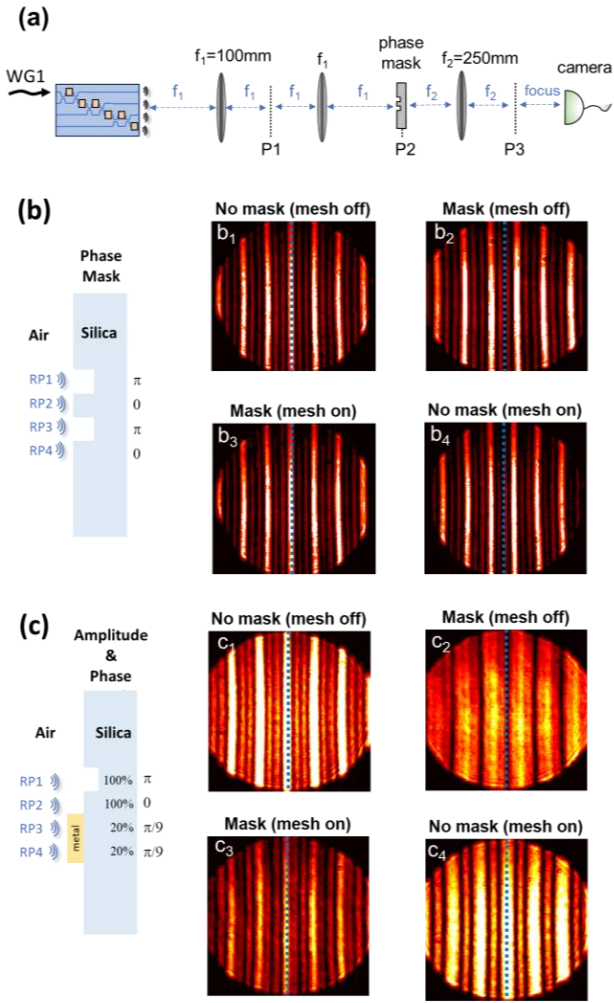

**Fig. 2.** Compensation of near-field amplitude and phase perturbations. (a) Optical setup employed to introduce intentional perturbations in the near-field generated by the output RPs of the mesh. An amplitude/phase mask is placed in the image plane P2 and the far field pattern is measured in plane P3 by a near-IR camera. In the inset the measured field in the image plane P2 is shown. (b) Compensation of pure phase perturbations: ($b_1$) reference far-field pattern with nominal RPs (same amplitudes and phases) and no mask in the system; ($b_2$) effect of a [$\pi$ 0 $\pi$ 0] phase mask; ($b_3$) recovery of the nominal far-field after mesh tuning; ($b_4$) far-field measured after mesh tuning if the phase mask is removed. (c) Compensation of amplitude and phase perturbations: ($c_1$) reference far-field pattern with nominal fields at the RPs and no mask in the system; ($c_2$) effect of a mask with [$\pi$ 0 $\pi/9$ $\pi/9$] phase profile and [1 1 0.2 0.2] amplitude profile. ($c_3$) recovery of the nominal far-field after mesh tuning; ($c_4$) far-field measured after mesh tuning if the phase mask is removed. Input optical power is increased in this experiment to maintain the clarity of pictures due to the loss introduced by mask, since this is a passive mesh (i.e., with no amplification).

Fig. 2($b_1$) shows the nominal ("perfect") far-field that is measured when the fields of the four RPs share the same amplitude and the same phase, and no other optical elements are introduced along the optical path (no obstacle in plane P2). In this condition, the far-field shows a periodic pattern of bright fringes along the direction parallel to the RP array. That direction is the horizontal axis of patterns in Fig. 2(b) and (c), while the vertical dashed blue line indicates the reference central axis of the camera. The main lobe is located at the center of the pattern in Fig. 2($b_1$), while the other ones (grating lobes) are spaced by 3.9-mm (0.69 degree) due to the $\sim 82\lambda$ spacing between array elements. Lower-power side-lobes are visible within each period, whose number depends on the number of RPs (see Supplementary Sec. 2).

To introduce perturbations equivalent to non-idealities in the field of the RPs, we insert an obstacle in the image plane P2. This obstacle is an engineered phase mask realized on a 0.5-mm thick silica substrate etched on one side at various depths in order to introduce a desired local phase shift of the transmitted light. The unetched surface is assumed as a reference phase, while an etch depth of 1.663 nm corresponds to a $\pi$ phase-shift. In the case reported in Fig. 2(b), the phase mask has a [$\pi$ 0 $\pi$ 0] phase profile and it is positioned in such a way that each phase shift contribution acts only on the field produced by an individual RP. As a result, in the far field (panel $b_2$) we observed a right-hand shift of the main lobe by half a period, corresponding to an angular deviation of 0.334 degree, in agreement with numerical simulations reported in Supplementary Fig. S3.

The perturbation introduced by the phase mask can be compensated by re-tuning the integrated mesh. The tuning procedure can be carried out automatically by measuring the optical power in the position where the main lobe of the far field should be nominally located (that is, in the center of the camera) and by driving the phase actuators in order to maximize it. In our experiment we employed a dithering based gradient descent algorithm [31], but alternative strategies such as evolutionary genetic algorithms [32] or machine learning [33], could be equivalently used. The far field pattern achieved at convergence of the tuning process is shown in Fig. 2($b_3$), demonstrating the recovery of the nominal beam observed without any phase mask ($b_1$). This means that the output phase profile at the RPs was set equal to the phase conjugate of the phase profile introduced by the mask. As a proof, we removed the phase mask while keeping the tuning state of the mesh and we observed the far field pattern of Fig. 2($b_4$), where the (opposite) left-hand half period shift is recorded due to the phase profile [$-\pi$ 0 $-\pi$ 0] at the output RPs of the mesh.

With respect to conventional optical phased arrays, integrated MZI meshes can also manipulate the amplitude of the light at the RPs. To examine this capability, we introduced in the light path a mask that introduces both amplitude and phase perturbations. The amplitude profile is realized by thermally evaporating and patterning onto the silica substrate of the mask a 40 nm thick chromium layer resulting in 20% transmission coefficient where it is present on the mask. A phase profile is added in the metal-free regions by locally etching the silica surface. The mask employed in the experiment of Fig. 2(c) introduces a [$\pi$ 0 $\pi/9$ $\pi/9$] phase shift and [1 1 0.2 0.2] amplitude transmission. After positioning the mask in the image plane P2, we recorded the perturbed far-field pattern of Fig. 2($c_2$) in agreement with simulations reported in Supplementary Fig. S4(a). In this scenario only compensating the phases of radiating elements is not enough to recover the ideal beam, but the amplitudes of the RPs must be modified as well. By using the same tuning procedure of the mesh employed in the case of Fig. 2(b), that is by maximizing the power at the center of the beam, we recorded the pattern of Fig. 2($c_3$), which is a faithful replica of the unperturbed pattern ($c_1$), yet with a lower power level. Indeed, a passive mesh cannot compensate for the average attenuation introduced by the mask; it can only redistribute the amplitude of the overall radiated power (though it can do so without any additional power loss). The far-field pattern observed (Fig. 2($c_4$)), when the mask is removed while maintaining the tuning status of the mesh, confirms this concept. From a comparison with simulations reported in Fig. S4(b), we can see that the mesh has modified the RPs phases as $[-\pi \ \ 0 \ \ -\pi/9 \ \ -\pi/9]$, which is the phase conjugate of the mask phase profile, and the RPs amplitudes are re-distributed in proportion to [0.2 0.2 1 1]; the result of this compensation of amplitude leads back to an overall uniformly distributed array after the mask attenuation is included.

It is worth pointing out that the automated tuning process implemented in our scheme does not aim at recovering the nominal far-field, but at finding the mesh configuration that optimizes a given metric in the far-field pattern. In the examples considered here, we used maximization of power along a specific direction (that is, in the center of the reference beam) as a metric. More generally, if the optimum pattern according to a certain metric is not known a-priori, this pattern can be found automatically in our approach [24].

## 4. Imaging through obstacles

The capability of a MZI mesh to operate as a programmable beam shaper can be exploited to transmit free space optical beams through obstacles that corrupt the beam front. Figure 3(a) shows the schematic of the free-space setup employed to demonstrate this concept. An obstacle is introduced in the optical path between the the MZI mesh and the camera focus plane at a far-field distance from the RPs and without any optical system in between.

As a first example the obstacle is a phase mask that is placed at a distance of 10 cm from the mesh output. Given the 3-degree beam width of the field emitted by each RPs, at this distance the overall beam pattern generated by the array has a full width of about 520 µm. As shown in Fig. 3(b) the phase mask is made of two transparent slits spaced by 300 µm that each introduce a $\pi$-shift across the phase front of the beam. The mesh is initially configured to maximize the light intensity at the center of the camera focus plane in the absence of the mask [Fig. 3($b_1$)] (which effectively means the mesh has configured itself to focus as well as it can on the focus plane of the camera). Upon the insertion of the mask the field pattern is altered and exhibits zero intensity at the center of the camera [Fig. 3($b_2$)]. The phase-front perturbation introduced by the phase mask can be counteracted by properly reconfiguring the MZI mesh, whose amplitude and phase output beams can be set to obtain maximum intensity at the center of the camera focus plane [Fig. 3($b_3$)]. Removing the phase mask, the far field of the beam generated by the mesh is different from ($b_1$) as shown in Fig. 3($b_4$). This means that the MZI mesh was configured indeed to establish the best communication channel in the presence of the obstacle [24].

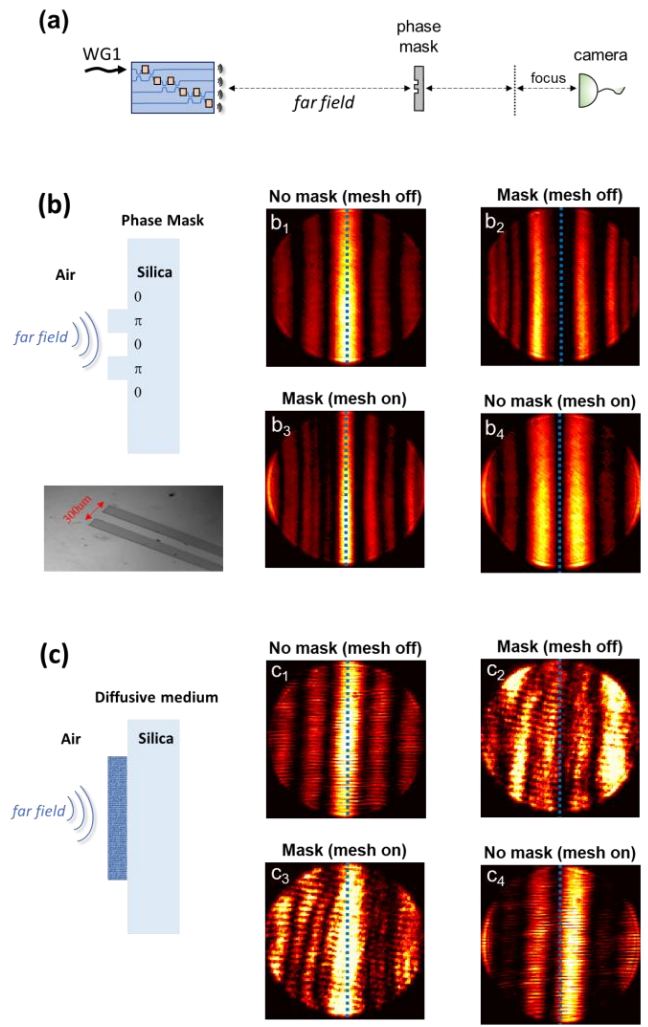

**Fig. 3.** Compensation of far-field perturbations. (a) Optical setup employed to introduce intentional perturbations in the far-field generated by the mesh. An obstacle is placed between the mesh and the camera with no other optical elements along the path. (b) Imaging through a phase mask: ($b_1$) reference far-field pattern with nominal RPs (sample amplitudes and phases) and no mask in the system; ($b_2$) effect of a [$\pi$ 0 $\pi$ 0] phase mask introduced in the far field; ($b_3$) recovery of the nominal far-field after mesh tuning; ($b_4$) far-field measured after mesh tuning if the phase mask is removed. (c) Imaging through a diffusive medium: ($c_1$) reference far-field pattern with nominal fields at the RPs and no obstacle in the system; ($c_2$) effect of scattering medium (scotch tape) introduced in the far field. ($c_3$) recovery of the nominal far-field after mesh tuning; ($c_4$) far-field measured after mesh tuning if the diffusive medium is removed.

Similar results can be achieved if the obstacle consists of a diffusive optical element. In this case the goodness in the recovery of the beam is only limited by the number of radiating elements of the mesh, which represent the available degrees of freedom that we have to compensate for the beam front distortion. In the experiment of Fig. 3(c) a plano-cylindrical lens coated with a scattering surface (a stripe of scotch tape) is inserted along the optical path. The presence of the scattering element perturbs the reference beam [Fig. 3($c_1$)] in such a way that the far field of the distorted beam has a null at the center of the camera [Fig. 3(c2)]. By retuning the mesh, the far field can be reshaped in order to maximize the power at the position of the detector [Fig. 3(c3)]. The new configuration of the mesh leads to an optimum beam phase front to compensate for the phase and amplitude perturbations caused by the scattering medium as much as allowed by the number and position of the RPs of the mesh. If the scattering medium is removed, the shape of the far field that is generated in this specific case does not have a maximum in the center [Fig. 3(c4)]. The perturbation from this distortion is not predictable without knowledge of the scattering medium, but here the beam has been reconstructed by automatically tuning the mesh using a feedback signal from the receiver and without any knowledge of the scattering medium structure.

### 5. Obstacle identification

The results reported in the previous sections demonstrate that no a-priori information on the obstacle inserted in the free-space light path is required for the tuning of the MZI mesh. This implies that the same tuning procedure can be used also to identify the profile of an unknown obstacle, provided that the system is suitably pre-calibrated.

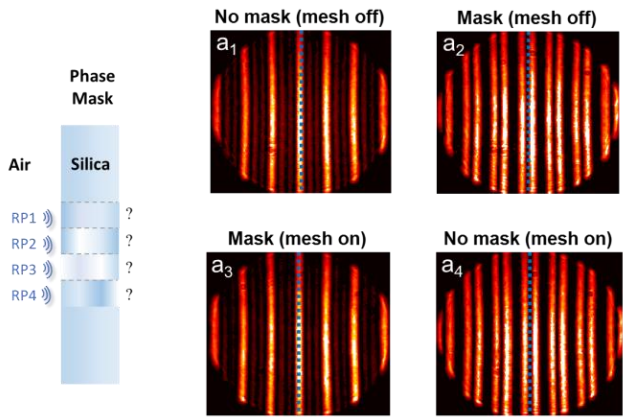

**Fig. 4.** Obstacle identification. (a1) reference far-field pattern with nominal RPs and no mask in the system; (b2) effect of an unknown mask introduced in the near field setup of Fig. 2; (b3) recovery of the nominal far-field after mesh tuning; (b4) far-field measured after mesh tuning if the mask is removed. This profile corresponds to a RP phase correction of [π π 0 0], thus indicating that the mask has a phase profile of [-π -π 0 0].

In the experiment shown in Fig 4 we introduce in the image plane P2 of the optical system of Fig. 2 a phase mask with four phase levels but with an unknown profile. After positioning of the mask along the light path, the measured far-field pattern changes from the reference shape of panel Fig. 4(a1) to the perturbed one of panel Fig. 4(a2). Compared with the unperturbed pattern, in each period two bright lobes now share the same power, while side lobes almost disappear.

In this configuration, no power is received by a detector placed in the middle of the far-field pattern. If the mesh is now reconfigured to maximize the power at the detector, the original unperturbed pattern is recovered, Fig. 4 (a3). Information on the phase correction introduced by the mesh on the optical field radiated at the output ports can be inferred from the voltages driving the phase actuators.

Comparing the set of the applied voltages with a preset lookup table, we estimated a phase correction of $[-\pi \; -\pi \; 0 \; 0]$. This result matches with the phase profile of the mask which was engineered to be $[\pi \; \pi \; 0 \; 0]$. By removing the phase mask, the far-field pattern radiated by the mesh appears as in panel (c4), whose intensity profile is the same as in (c2), as expected from the fact that a phase shift of $+\pi$ (phase mask) or $-\pi$ (mesh) in the array produces the same effect on the radiated pattern.

### 6. Conclusion

We demonstrated the use of a programmable mesh of integrated MZIs to effectively manipulate the properties of FSO beams. The mesh can be automatically configured to shape FSO beams with an ideal pattern also with non-perfect optical antennas and to generate the best FSO beam (given the number of RPs of the mesh) that can propagate through deterministic obstacles (amplitude and phase masks) or random media (scattering medium). Given the number of RPs of the mesh, representing the (complex-valued) degrees of freedom of the system, the mesh is configured to establish the best communication channel with a receiver in the presence of the obstacle [24]. If the obstacle is a pure phase mask, the mesh automatically generates the conjugate of the phase profile of the mask in the plane of the obstacle; in the case of additional amplitude screening given by the obstacle, the mesh also redistributes the power across the beam front in order to generate a uniform amplitude profile after the obstacle. This is a significant advantage compared to SLMs, where amplitude modulation is achieved by introducing optical loss [7]. Multiple output diagonal meshes can guarantee multiple orthogonal diffraction limited free space channels between two integrated meshes [24].

The automated strategies used for the calibration of the MZI mesh advantageously exploit the implementation of local feedback loops in each MZI stage, which have been demonstrated effective in previous reports [13]; the benefit of this approach is the much lower complexity with respect to global multivariable optimization techniques that are needed in other devices employed for the manipulation of FSO beams, such SLMs. This would also allow the possibility of dynamically compensating and tracking time-varying changes in the FSO link, resulting, for instance, from the presence of particles or turbulence, so as to maintain the optimum communication link.

The procedure employed for the tuning of the mesh does not need any information on the nature of the obstacle. Therefore, after proper calibration the system, it can be effectively used for the identification of an unknown obstacle inserted in the FSO path. Since the MZI mesh can generate any relative amplitude and phase at its RPs, obstacles with arbitrary amplitude and phase profile can be recognized, provided that the number of RPs is equal or larger than the degrees of freedom of the obstacle. In other words, the mesh could be used as a reader for multilevel amplitude and phase bar codes.

Finally, the implementation of the MZI mesh on a standard silicon photonic platform makes it straightforwardly integrable with many available devices performing optical functions like wavelength filtering, (de)multiplexing and fast time-domain modulations, thus enabling implementation of more advanced all-optical processing of FSO beams.

**Funding sources and acknowledgments.** This work was supported by the European Commission through the H2020

project Super-Pixels (grant 688172). DABM acknowledges support from the Air Force Office of Scientific Research (AFOSR) under award number FA9550-17-1-0002.

**Acknowledgments**. This work was partially performed at Polifab, the micro- and nanofabrication facility of Politecnico di Milano (https://www.polifab.polimi.it/). We are grateful to F. Zanetto and M. Petrini for support in the design of the electronic control systems.

**Disclosures.** The authors declare no conflicts of interest.

# Automated manipulation of free space optical beams with integrated silicon photonic meshes

**Maziyar Milanizadeh[1], Fabio Toso[1], Giorgio Ferrari[1], Tigers Jonuzi[1,+], David A. B. Miller[2], Andrea Melloni[1], Francesco Morichetti[1,*]**

[1]*Dipartimento di Elettronica, Informazione e Bioingegneria - Politecnico di Milano, Milano, 20133 Italy*
[2]*Ginzton Laboratory, Stanford University, Spilker Building, Stanford, CA 94305, USA*
*\*Corresponding author: francesco.morichetti@polimi.it*
[+]*now with VLC Photonics, Universidad Politécnica de Valencia, 46022 Valencia, Spain*

## S1. Imaging the field of the radiation ports (RPs)

To introduce engineered perturbations in the plane of the radiation ports of the mesh, which are spaced by 127 μm, we fabricated a phase mask by etching depths into the surface of a 2-inch silica wafer. Figure S1(a) shows the picture of this mask including different phase patterns. Due to the size of the phase mask and to the assembly of the silicon chip wired to the electronic PCB for electrical I/Os [Fig. S1(b)], the mask cannot be placed directly on top of the chip. To overcome this problem, the "4f" optical relay setup in Fig. 2(c) is used, where two lenses with focal length $f_1$ = 100mm create a one-to-one replica of the field at the RPs in a plane P2 at a suitable distance from the chip. This approach also gives an intermediate Fourier plane P1.

For calibration of the setup a NIR camera was used to observe the field pattern in plane P2. Since the camera pixels have 20μm pitch and the objective lens ($f$ =35mm) used in front of the camera has a 6.5 times demagnification at the nearest focal distance, we implemented a magnification system, using a single positive lens with 50 mm focal length, as in Fig. S1(c). Figure S1(d) presents the magnified picture taken from plane P2 showing the replicated image of the four RPs fields with a small aberration effect due to the simple single-lens magnification approach. The overall magnification of the imaging system is 0.5X, resulting in a spacing of the imaged RP fields by about 60 μm, corresponding to 3 pixels of the NIR camera.

## 2. Numerical simulation of four-element array

In this section we provide some details on the radiation pattern generated in the far field by a linear array of $N$ = 4 uniformly spaced grating couplers spaced by $d$ = 127 μm. A picture of the actual array is shown in Fig. 2(a). Both in simulation and experiments the array is excited uniformly – that is, the optical field at each radiator has same amplitude and phase.

As a simulation platform for the phased array system, we used the *Sensor array analyzer* toolbox from MathWorks [S1]. An optical beam with a gaussian shape and a 3-degree beamwidth is assumed for the field emitted by each array element, as derived by EM simulations of the radiation pattern of the individual grating coupler. The simulated far-field pattern of the 4-element array is shown in Fig. S2(b) along the Azimuth direction (parallel to the array) at an elevation angle of 0 degree. Since the array elements are spaced by many wavelengths (about 82$\lambda$), the array function exhibits a sequence of transmission peaks located at angles

$$\theta = 90° - \cos^{-1}\left(M\frac{\lambda}{d}\right) \quad (1)$$

where $\theta$ is the azimuth angle and $M$ is an integer. The main lobe is associated with $M$ = 0, while the replicas at $M$ ≠0 are referred to as grating lobes. Within each period there are $N$-2 two side lobes and $N$-1 zeros. In our case, the angular spacing between the main lobe and the first grating lobe is 0.69 degree. The maxima of the grating lobes decrease with increasing $\theta$ according to an envelope function given by the radiation pattern of an individual radiating element.

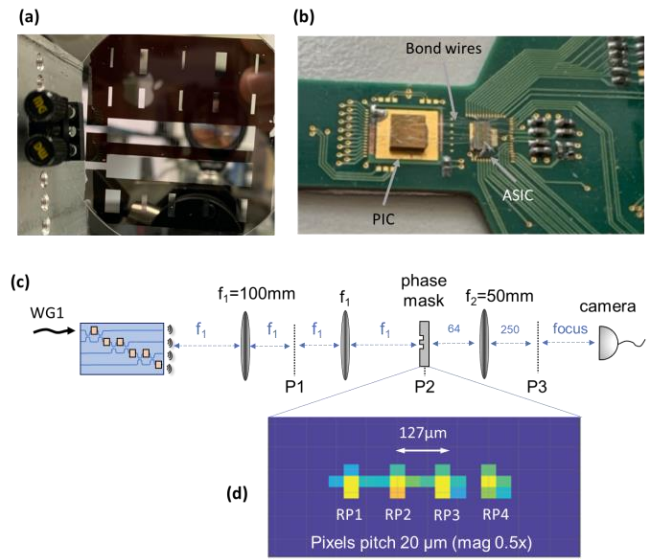

**FIG. S1.** (a) Photograph of the phase mask employed to introduce intentional perturbation of the field at the RPs of the mesh, (b) Photograph of the silicon chip assembled onto the electronic PCB; (c) Schematic of the experimental setup employed to create a one-to-one replica of the field at the RPs in plane P2. (d) Image of the four RPs fields measured in plane P2.

Figure S2(c) shows the 2D simulation of the far field pattern within an azimuth range of ±2.2 degree (vertical axis) and an elevation range of ±4 degree (horizontal axis). A very good agreement is found with the experimental result of Fig. S2(d), showing the portion of the far field pattern captured by the NIR camera in the same excitation condition.

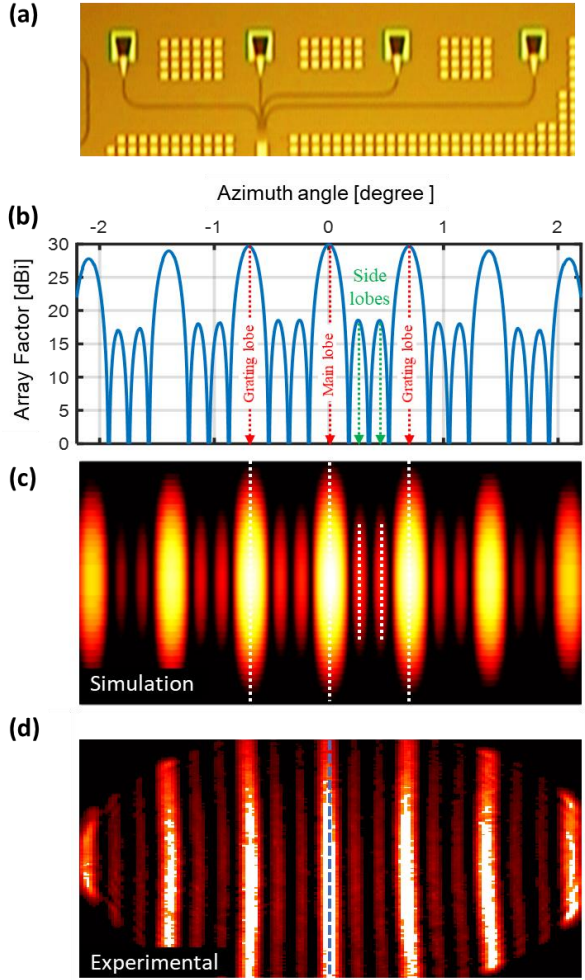

**Figure S2**: (a) Top-view photomicrograph of the uniform linear array of optical radiation elements consisting of vertical grating couplers spaced by 127 μm; (b) Simulated far-field pattern of the 4-element array in the azimuth direction (parallel to the array) at an elevation angle of 0 degree. (c) simulated and (d) measured 2D field pattern within an azimuth range of ±2.2 degree (vertical axis) and an elevation range of ±4 degree (horizontal axis). The vertical dashed blue line indicates the reference central axis of the camera.

Specific simulations were performed to get a better understanding of the behavior of the mesh in controlling the amplitude and the phase of the optical field at the RPs in the different experiments reported in the main text.

Figure S3(a) shows the simulated far-field pattern when a phase profile of $[\pi\ 0\ \pi\ 0]$ is applied to the field emitted by the four RPs, this phase shift being the same as the one given by the phase mask used in the experiment of Fig. 2. For a better comparison between the simulated and the measured results, Fig. 2($b_2$) of the main text is here reported in panel (b). A very good agreement is observed, the field pattern being shifted by half period to the right-hand side, that is a 0.334-degree steering. Introducing a $[-\pi\ 0\ -\pi\ 0]$ phase shift in the simulation, corresponding to the phase profile needed to compensate for the effect of the phase mask, we obtain the same pattern as in Fig. S3(a), which is due to a half-period shift in the reverse direction (left-hand steering). This result indeed matches the beam pattern measured in Fig. 2($b_4$).

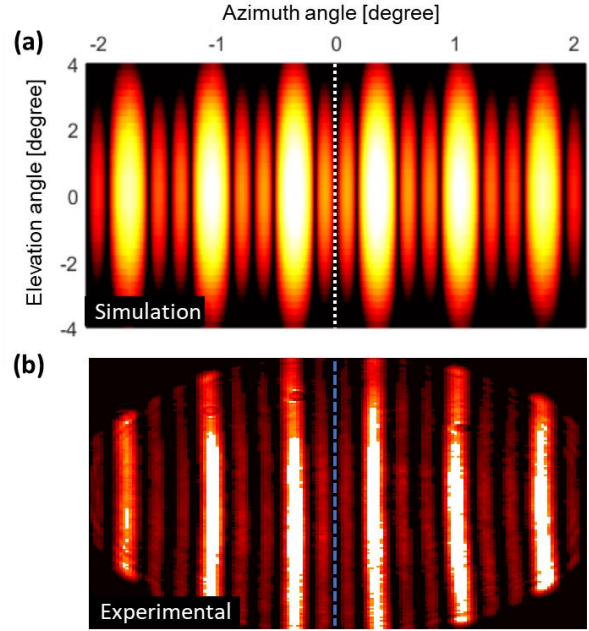

**Figure S3:** (a) simulated and (measured) 2D far-field pattern of the 4-element array when a phase profile of $[\pi\ 0\ \pi\ 0]$ is applied to the field emitted by the four RPs. The vertical dashed blue line indicates the reference central axis of the camera.

Simulation analyses were also carried out to evaluate the effects of complex obstacles introducing phase and amplitude perturbations. To emulate the experimental conditions of Fig. 2(c), where an amplitude and phase mask is inserted in plane P1, we applied a phase profile of $[\pi\ 0\ \pi/9\ \pi/9]$ to the array elements of Fig. S1(a), while considering an amplitude distribution proportional to [1 1 0.2 0.2]. Figure S4(a) shows that the simulated far-field pattern obtained in this condition matches the measurement of Fig. 2($c_2$). To counteract the effect of this perturbation, the elements of the array should be excited with a phase profile of $[-\pi\ 0\ -\pi/9\ -\pi/9]$ and an amplitude modulation proportional to [0.2 0.2 1 1] (which is obtained through redistribution of power as [0.33 0.33 1.66 1.66] in RPs). The simulated far-field pattern of the array operated in this configuration is shown in Fig. S4(b). This result is in a very good agreement with the measured far-field pattern of Fig. 2($c_4$), which is produced by the mesh after the tuning procedure targeted to compensate for effects of phase mask.

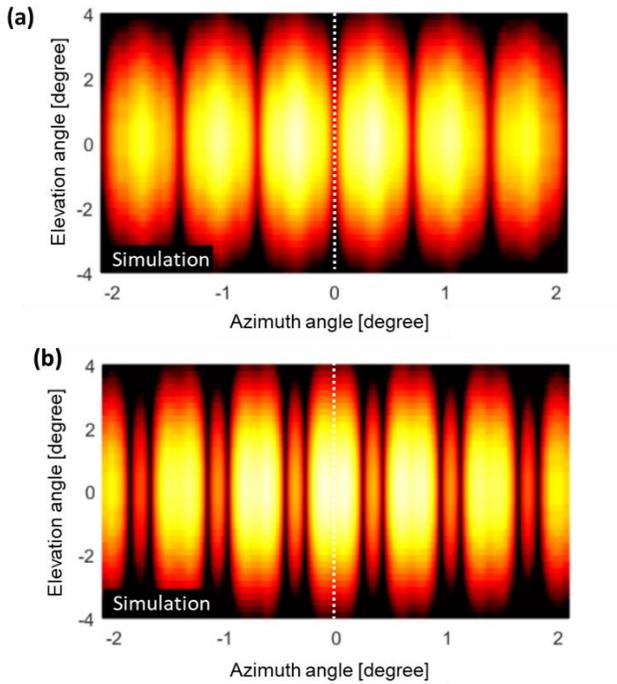

**FIG. S4:** (a) Simulated 2D far-field pattern of the 4-element array when a phase profile of [π 0 π/9 π/9] and an amplitude modulation proportional to [1 1 0.2 0.2] are applied to the field emitted by the four RPs. (b) Simulated 2D far-field pattern of the 4-element array when a phase profile of [-π 0 -π/9 -π/9] and an amplitude modulation proportional to [0.2 0.2 1 1] are applied to the field emitted by the four RPs.